\documentclass[conference]{IEEEtran}
\IEEEoverridecommandlockouts
\usepackage{amsmath,amssymb,amsfonts}
\usepackage{algorithmic}
\usepackage{graphicx}
\usepackage{textcomp}
\usepackage{booktabs}
\usepackage{xcolor}
\usepackage{pifont}
\usepackage{url}
\usepackage{hyperref} 
\usepackage{subcaption}
\newcommand{\xmark}{\ding{55}}
\renewcommand{\checkmark}{\ding{51}}
\usepackage[colorinlistoftodos]{todonotes}
\def\BibTeX{{\rm B\kern-.05em{\sc i\kern-.025em b}\kern-.08em
    T\kern-.1667em\lower.7ex\hbox{E}\kern-.125emX}}

\usepackage[backend=biber, style=ieee, maxnames=3, minnames=1]{biblatex}  % up to maxnames for full display; up to minnames for trimmed display
\usepackage{setspace} % for line spacing within entries
\makeatletter

\def\ps@IEEEtitlepagestyle{%
    \def\@oddfoot{\mycopyrightnotice}%
    \def\@evenfoot{}%
}
\def\mycopyrightnotice{%
    {\footnotesize  DOI 10.1109/HOTI66940.2025.00021 \textcopyright2025 IEEE\hfill}
    \gdef\mycopyrightnotice{}
}
\makeatletter
\newcommand*\titleheader[1]{\gdef\@titleheader{#1}}
\AtBeginDocument{%
  \let\st@red@title\@title
  \def\@title{%
    \bgroup\small\large\centering\@titleheader\par\egroup
    \vskip1.5em\st@red@title}
}
\makeatother

\title{Accelerating Frontier MoE Training with 3D Integrated Optics}

\titleheader{2025 IEEE Symposium on High-Performance Interconnects (HotI)}

\begin{document}

% \title{Accelerating Frontier MoE Training with 3D Integrated Optics}

\author{\IEEEauthorblockN{Mikhail Bernadskiy, Peter Carson, Thomas Graham, Taylor Groves, Ho John Lee, Eric Yeh}
\IEEEauthorblockA{\textit{Lightmatter} \\
}

}

\maketitle
\begin{abstract}
The unabated growth in AI workload demands is driving the need for concerted advances in compute, memory, and interconnect performance. As traditional semiconductor scaling slows, high-speed interconnects have emerged as the new scaling engine, enabling the creation of larger logical GPUs by linking many GPUs into a single, low-latency, high-bandwidth compute domain.  While initial scale-up fabrics leveraged copper interconnects for their power and cost advantages, the maximum reach of passive electrical interconnects (approximately 1 meter) effectively limits the scale-up domain to within a single rack. The advent of 3D-stacked optics and logic offers a transformative, power-efficient scale-up solution for connecting hundreds of GPU packages (thousands of GPUs) across multiple data center racks.

This work explores the design tradeoffs of scale-up technologies and demonstrates how frontier LLMs necessitate novel photonic solutions to achieve aggressive power and performance targets. We model the benefits of 3D CPO (Passage) enabled GPUs and switches within the scale-up domain when training Frontier Mixture of Experts (MoE) models exceeding one trillion parameters. Our results show that the substantial increases in bandwidth and radix enabled by 3D CPO allow for an 8X increase in scale-up capability. This affords new opportunities for multi-dimensional parallelism within the scale-up domain and results in a 2.7X reduction in time-to-train, unlocking  unprecedented model scaling.

\end{abstract}

%\begin{IEEEkeywords}

%\end{IEEEkeywords}

\section{Introduction}
The race to build larger, more sophisticated AI models is pushing the limits of existing infrastructure.
At the chip and package level, GPUs are constrained by shoreline, yields and power.  These challenges have led to the development of large high-bandwidth, low-latency scale-up pods. These pods effectively combine hundreds of GPUs into a single logical GPU to facilitate a variety of parallelism strategies (e.g. Data, Tensor, Expert) for large AI models.  Approaches like Mixture of Experts (MoE)~\cite{shazeer2017outrageouslylargeneuralnetworks} have pushed scale-up networks to their limits due to copper reach (1 meter), which constrains the number of GPUs that can be connected within a single network hop.

With MoEs, an ensemble of specialized sub-networks work together through sparse activations to increase model capacity without significantly increasing computational requirements.  The output of the selected experts are combined to create the final result.  MoE allows models to scale and learn more nuanced representations, but adds additional communications overhead as each set of top-k experts use costly all-to-all operations.  Studies have shown that the communication involved in expert parallelism can account for 47\% of the forward pass latency, even when utilizing a high-bandwidth scale-up interconnect (7200 Gbps)~\cite{li2025speculativemoecommunicationefficient}.  Larger scale-up domains directly translate into the capability to deploy a larger number of experts and improve model performance.
  
This paper explores the limitations of current approaches and presents a paradigm shift: the transition from copper to integrated 3D photonics in order to create a more scalable and efficient high-bandwidth domain across racks in the data center.  In this work, we show: 
\begin{itemize}
    \item   Passage has the unique combination of bandwidth density, port count, reach and energy efficiency to enable multiple generations of innovation in AI infrastructure bringing a \textbf{8X increase to scale-up pod bandwidth using half the energy of conventional CPO}.
    \item Comparisons of system design tradeoffs using electrical, pluggable optics modules, CPO and 3D integrated optics, showing impressive advantages in area and density  resulting in a \textbf{6X reduction in package area expansion compared to CPO}.
    \item Application benefit of an expanded scale-up domain for LLM training, demonstrating \textbf{2.7X} speedup in training time compared to electrical designs.
\end{itemize}

We begin with a background on LLM training, scale-up networking and motivation for 3D integrated optics (3D).  We then provide an overview of the Passage platform, highlighting the benefits of Passage in terms of bandwidth density, energy efficiency. (Section~\ref{sec:Passage}).  In Section~\ref{sec:System} we examine how system architects could construct a scale-up domain out of different technologies (LPO, CPO and 3D optics) highlighting the tradeoffs of each approach.  We use these systems designs to model performance of frontier LLM training using Mixture of Experts in Sections~\ref{sec:App} and ~\ref{sec:Results}.  Finally, we provide conclusions and discuss future directions for innovation.

\section{Background}
\label{sec:background}

\subsection{LLM Training}
Since 2017, transformer-based language models have steadily increased in size, with higher parameter counts enabling increasingly powerful model capabilities. The original 65M parameter transformer~\cite{vaswani2023attentionneed} was trained on a single 8-GPU node, while recent frontier models have on the order of 1 trillion parameters and are trained on datacenter-scale clusters~\cite{semigpt4arch}~\cite{llama4herd}.
Models are trained using gradient descent methods, each step requiring a forward pass on a batch of training input, evaluation of a loss metric, and a backward pass computing loss gradients and parameter updates.
The compute and memory requirements for training a transformer are dominated by the attention block and the feed-forward network (FFN) in each layer, which are mostly matrix multiplication operations. Tensor parallelism~\cite{shoeybi2020megatronlmtrainingmultibillionparameter} is commonly used to distribute a single layer across multiple GPUs to speed up compute throughput and increase the memory available for model parameters, activations, and optimizer state. 

\begin{figure}[t]
    \centering
    \begin{subfigure}[b]{0.35\textwidth}
        \centering
        \includegraphics[width=\textwidth]{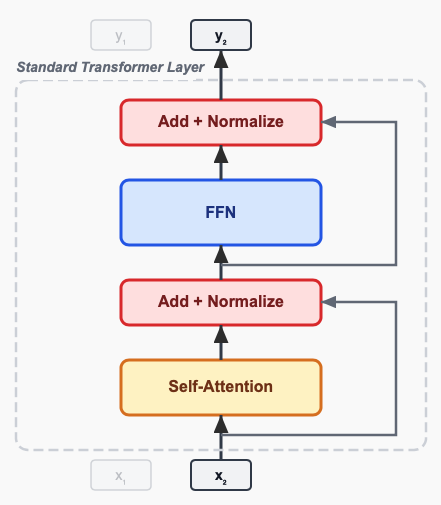}
        \caption{Dense Model}
        \label{fig:dense-model}
    \end{subfigure}
    \hfill
    \begin{subfigure}[b]{0.35\textwidth}
        \centering
        \includegraphics[width=\textwidth]{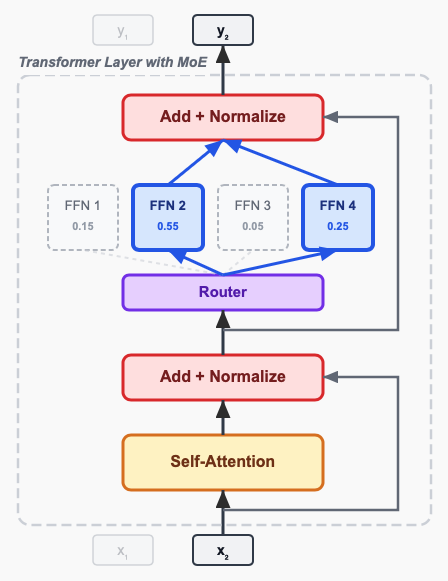}
        \caption{Mixture of Experts (MoE) Model}
        \label{fig:moe-model}
    \end{subfigure}
    \caption{Transformer architectures: (a) Dense model with self-attention and FFN. (b) Sparse MoE with top-k=2 routing selecting experts E\textsubscript{2} and E\textsubscript{4}  based on highest scores shown for token x\textsubscript{2}.} 
    %\caption{Comparison of transformer architectures. (a) Standard dense transformer block with self-attention and feed-forward networks. (b) MoE sparse model with top-k=2 routing, where the router selects experts E\textsubscript{2} and E\textsubscript{4} (the two highest probability %experts) for token x\textsubscript{2}. The probability is shown for each expert.}
    \label{fig:moe_layer}
\end{figure}

In sparse Mixture of Experts (MoE) transformer models~\cite{shazeer2017outrageouslylargeneuralnetworks}~\cite{jiang2024mixtralexperts}, the FFN layer in the original (now referred to as "dense") transformer is replaced by multiple "experts" (Figure \ref{fig:moe_layer}), which are frequently identical to the original FFN network, and a small additional routing network selects which and how many alternatives should be activated for each token. This enables a larger, potentially more expressive model size at a given amount of compute, and the opportunity to "upcycle"~\cite{he2024upcyclinglargelanguagemodels}
previously trained dense models into larger sparse MoE models. The compute and data patterns mostly remain as before, but with an additional pattern for routing tokens to selected experts within an MoE layer ("expert parallelism"). 
% \begin{figure*}[h]
%        \centering
%    \includegraphics[trim={0.5cm 0 0 0},clip,width=0.95\linewidth]{figs/fedus et al 2022 cs336.png}
%    \caption{Correspondence of feedforward layers in dense vs sparse transformer model  \cite{fedus2022switchtransformersscalingtrillion}}
%    \label{fig:moe_layer}
%\end{figure*}

\subsection{Scale-up Networking}
Historically, the GPU interconnect bandwidth was limited by PCIe, and inter-GPU connectivity was limited by the network interface card. The advent of NVLink 1.0 for the Pascal generation of GPUs~\cite{foley2017ultra} allowed for a limited number of GPUs to create a high-bandwidth scale-up domain at 5X the PCIe bandwidth.  This was a massive leap forward in bandwidth and effectively enabled the multi-GPU tensor parallelism prominent in modern training.

\begin{table}
\centering
\caption{A comparison of scale-up vs scale-out networks}
\label{tab:scaleup}
 \resizebox{\columnwidth}{!}{%  % scales to column width
\begin{tabular}{| l | l | l | l | l |}
\hline
\textbf{Network Type} & \textbf{no. GPUs} & \textbf{latency} & \textbf{Tbps/GPU} & \textbf{Energy}\\
\hline
Scale-out & \textgreater 100k& 2-10 $\mu$s & 1.6 Tb/s& 16 pJ/bit~\cite{naddod800}\\
\hline
Scale-up & \textless 1024& 100-250 ns& \textgreater 12.8 Tb/s& \textless 5 pJ/bit\\
\hline

\end{tabular}
}
\end{table}

\begin{figure}[ht!]
    \centering
    \includegraphics[trim={0 0 0 0},clip,width=0.95\linewidth]{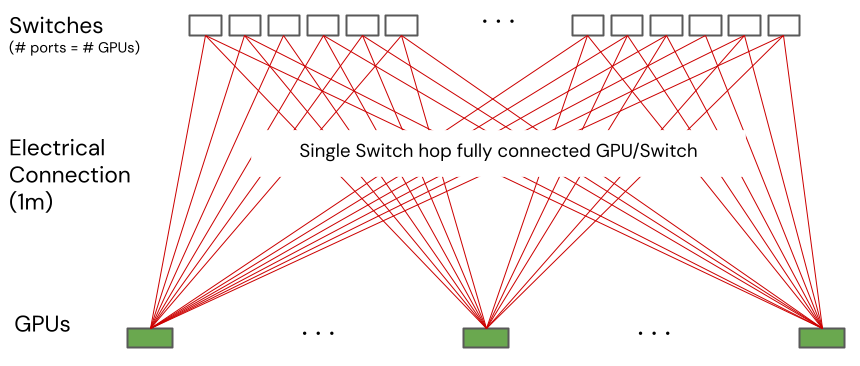}
    \caption{Single-layer Switch electrical scale-up topology. A single layer of switches (top) is connected to every GPU (bottom) in the pod (only three GPU-to-switch connections shown for brevity).  This provides full bandwidth connectivity between any two GPUs in the pod over multiple rails.}
    \label{fig:ElecPod}
\end{figure}

As the number of GPUs increased, switches were incorporated into the design to facilitate the increased bandwidth between a larger number of accelerators.
Scale-up topologies generally follow one of two approaches. The first is a multi-dimensional torus such as those deployed by the Google TPU network~\cite{jouppi2023tpu}.  A torus network provides efficient scaling, but incurs a large network diameter.  This is fine for deterministic ring-based collective algorithms, such as those employed by tensor parallelism or pipeline parallelism, but can experience congestion and delay for more general traffic patterns, such as expert parallelism with a non-deterministic set of experts.  The other commonly deployed topology is a single layer of switching (SLS), which uses multiple GPU rails to switch connections.  This is inherently a low-latency network with deterministic routing and performance.  This allows full bandwidth between any pair of GPUs, but the scale of the network is limited to the number of ports on the switch (i.e. a 512 port switch can support at most 512 GPUs -- one port per GPU).  Because of these characteristics, we focus on the SLS topology for the purposes of this paper.

The size of the GPU scale-up domain has continued to increase over time.  While the Nvidia Blackwell DGX pod supported 72 GPUs in 2024, 144 radix scale-up switches have been announced  to support 144 GPU packages in 2027~\cite{gtcroadmap}. The limiting factor in scaling the pod beyond 144 packages has been the reliance on copper and electrical networking.  As was stated in Nvidia GTC 2024, using pluggable optics modules would have required 20 kW, just to drive the NVLink spine.  This is a considerable amount of power, given a 120 kW rack budget~\cite{gtc2024}.  While electrical networking provides benefits in terms of simplicity and energy efficiency, the reach limitations at high SerDes data rates mean an electrically connected GPU pod is effectively limited to one or two racks.  For some, power is a secondary concern compared to the potential benefits of a larger scale-up domain.  Huawei has announced a fully optical scale-up domain that supports up to 384 AI accelerators in their Cloud Matrix design~\cite{CloudMatrix} with over a petabit per second of bandwidth for a single pod.  To construct this, they leverage pluggable optical modules, which we discuss further in Section~\ref{sec:ExistingOptics}.  As the size and bandwidth demands of the scale-up network continue to increase over time, 3D integrated optics provide the ultimate solution, with the bandwidth density and energy efficiency of electrical networks but longer reach.

\subsection{Motivation for 3D Integrated Optics}
\subsubsection{Package Growth and Shoreline Limitations}
Slowdowns to Moore's Law and Dennard Scaling have necessitated larger packages and increased power to deliver next-generation GPU performance. As packages grow, computational capability grows proportional to the area while the I/O is limited to the perimeter of the GPU.  To complicate matters, large portions of the shoreline are reserved for HBM, which require short trace lengths for signal integrity.

\begin{figure}
    \centering
    \includegraphics[width=0.95\linewidth]{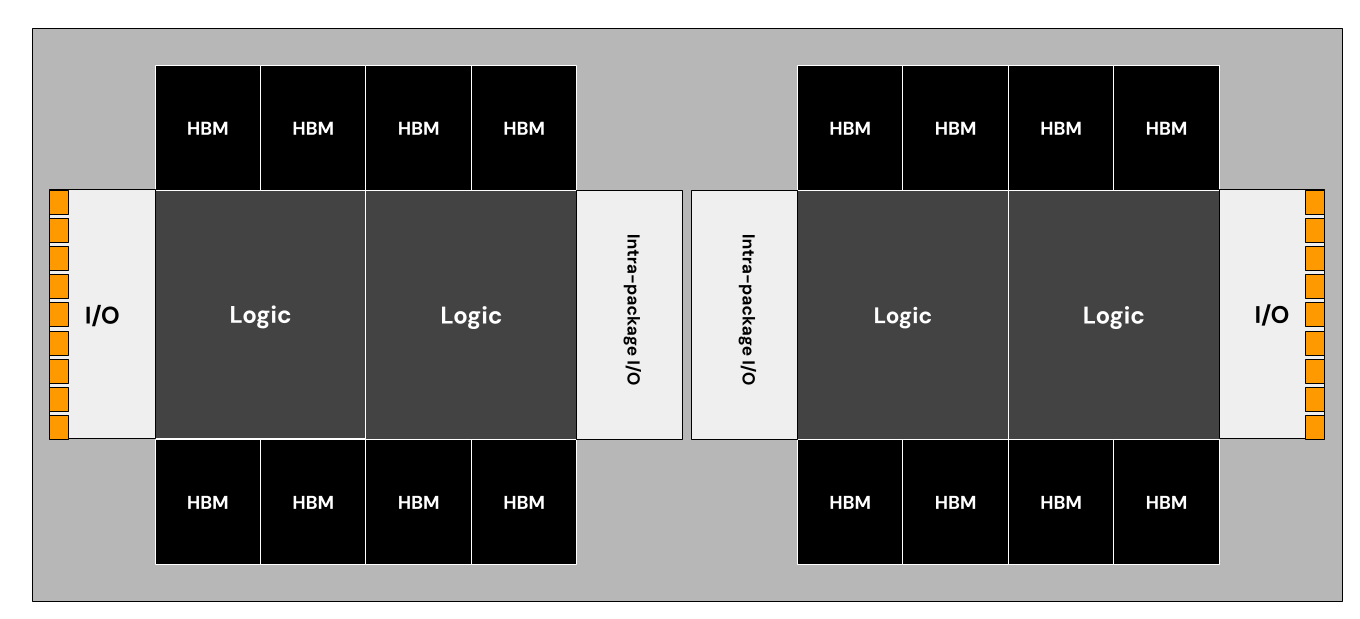}
    \caption{A GPU package in a 4 × 1 reticle configuration. Four logic reticles surrounded by HBM stacks on the north and south side in black, intra-package I/O in the middle and inter-package I/O on the east and west side.  SerDes Shoreline is highlighted in orange.}
    \label{fig:4x1}
\end{figure}

Figure~\ref{fig:4x1} shows an example of a GPU package where four logic chips, 16 stacks of HBM and I/O dies are all placed on a substrate.  Both I/O and HBM compete for the shoreline of the chip, leaving only the east and west available for scale-up bandwidth. 
For each I/O die, the bandwidth is limited by the number of SerDes macros that can fit along an edge. Doubling the bandwidth of these SerDes from 224 Gb/s to 448 Gb/s creates signal integrity challenges which require sophisticated equalization and increased power. 

\subsubsection{Electrical Reach Limitations}
As speeds of SerDes increase, electrical reach of the signal is reduced.  At 224 Gb/s the reach of passive Direct Attached Copper (DAC) is approximately 1 m, and at 448 Gb/s the reach is expected to be tens of centimeters.  To reduce the insertion loss, high-speed electrical solutions are moving towards co-packaged copper, and flyover cables to bypass lossy PCB traces.
For longer distances retimers must be deployed, which increases power.  The short reach of copper means GPUs and switches must be densely configured within a single rack.  This creates rack-level power challenges and shifts costs to cooling and infrastructure.  Current electrical systems are challenged to move beyond 72 GPU packages within a single pod.

\subsubsection{Challenges for Existing Optical Solutions}
\label{sec:ExistingOptics}
Optics have been deployed successfully for decades for applications where the distance between endpoints surpasses the capability of electrical transmission.  This includes long-haul (across continents), between 
datacenters (metro-regional) and within datacenters (hundreds of meters).  For each environment, the technology is optimized for differing criteria.  Since this work is focused on scale-up networks, we will discuss only the intra-datacenter application of optics.

The hurdles to deploying optics broadly within the datacenter have been cost, reliability and energy efficiency.
While optics cost more than passive copper solutions, the benefits of expanded reach add the potential for increased GPUs in the scale-up domain and faster time to solution.  We demonstrate this benefit in Section~\ref{sec:Results}.
Optics typically use lasers to power the transmission.  Lasers add cost, power, and can be temperature sensitive, failing at higher rates compared to copper connections.  Laser solutions must have fault tolerance and field-replaceable features baked into the design when operating at datacenter scale.  Also, the fiber connections are sensitive to contaminants or dust, making replacement a potential source of failure as well.  All of these components must be tested to ensure they are known-good before incorporating into a system.
 
 Optics enable disaggregation of the scale-up pod, creating an opportunity for power and cooling savings at the rack level, but given the massive amount of scale-up bandwidth (order of magnitude greater than scale-out), optics must be incredibly power efficient to fit within the GPU package and tray power budgets. At 5 pJ/bit, optics is effectively at parity with passive copper based solutions~\cite{shivnaraine202111, guo2022112} and 14.4 Tb/s of scale-up bandwidth results in 72 W of power per GPU.  At 20 pJ/bit this increases to over 288 W per GPU and reduces power available to computation.  20 pJ/bit effectively makes higher levels of scale-up bandwidth infeasible.  Energy efficiency of the scale-up network is paramount. 

\begin{table}[h]
    \resizebox{\columnwidth}{!}{%  % scales to column width
    \centering
    \begin{tabular}{|c|c|c|c|}
        \hline
        & Optical Module & LPO incl. & 2/2.5D CPO incl.\\
        & incl. Host SerDes & Host SerDes & Host SerDes \\ \hline
        Bandwidth & \xmark & \xmark & Medium \\ 
        Density & & & \\ \hline
        Energy & 21 pJ/bit & 13 pJ/bit & 12 pJ/bit \\ 
        Efficiency & \cite{naddod800} & \cite{OIFOFC, EoptolinkLPO, synopsysLPO} & \cite{BaillyHotChips} \\ \hline
        Latency & High (Retimed) & Medium & Low\\ \hline
        & & & with external \\
        Serviceability & $\checkmark$ & $\checkmark$ &  laser and \\
        & & & plug. coupler\\ \hline
        Std. Mechanical & & & \\ 
        Form Factor & $\checkmark$ & $\checkmark$ & \xmark \\ \hline
        Link  & $\checkmark$ & Co-design & Co-design \\ 
        Interoperability & & with host & with host \\ \hline
        HVM & $\checkmark$ & $\checkmark$ & 2026 \\ \hline
    \end{tabular}
    }
    \vspace{1em}
    \caption{Comparison of the key qualities associated with legacy optical technologies.  Energy efficiency assumes 5 pJ/bit for LR class 112 Gb/s PAM-4 SerDes with DSP on the host ~\cite{shivnaraine202111, guo2022112} (e.g. GPU or switch) plus 16 pJ/bit for optical module, 8 pJ/bit for DR8 LPO and 7 pJ/bit for 2.5D CPO and laser.}
    \label{tab:legacy}
    
\end{table}

\paragraph{Optical Modules}

Typical pluggable optical modules (e.g., OSFP) often integrate power-hungry DSPs and retimers to overcome host-to-module signal loss, resulting in high aggregate power (e.g., 21 pJ/bit) and large form factors ($>2000$ sqmm). While easily field-replaceable and interoperable across platforms, their inherent power consumption and significant area footprint limit density.

\paragraph{Linear Pluggable Optics}

LPO transceivers are an optimization of conventional pluggable optics modules such that the DSP is removed from the module itself. It is a linear drive in the sense that the signal from one host to another host device does not require a retimer or incur the extra power and performance overheads.  The expectation is that an LPO module is approximately 25-50\% more power efficient than a conventional pluggable module~\cite{synopsysLPO, OIFOFC}.  This creates a reliance on the host-side interface to do the heavy lifting and drive the signal without retimers.  Therefore host SerDes in an LPO-based system are expected to rely on DSPs and be in the range of 4.5-6 pJ/bit \cite{shivnaraine202111, guo2022112}.  LPO solutions must be co-designed in consideration of the host platform capabilities and link budget specific to a given end device (GPU, CPU, Switch, etc).  In some cases LPO optics utilize flyover cables between the host and the module to reduce losses further, but this adds expense and complexity.  LPOs still leverage large form factors (e.g. OSFP-XD) resulting in low bandwidth density compared to integrated optics.  As data rates and the number of channels per module increase the modules may require cold-plate cooling.

\paragraph{Co-packaged Optics}

Co-packaged optics describes the process of taking the optical transceiver and moving it onto the same package as the device it is supporting (typically a processor or switch).  It is compelling because you reduce the distance traveled electrically over a high loss medium such as a PCB, and decrease latency compared to a pluggable optical module.  When discussing co-packaged optics it is helpful to distinguish between the approach taken to build the Optical Engine and how it's integrated into the host package.  In both contexts, the concepts of 2D, 2.5D and 3D design can apply.

The integration with the host may be 2D or 2.5D.  A good overview of this topic has been provided by Lee, Nedovic, Greer and Gray~\cite{beyond}.  In both 2D and 2.5D packaging, the host chip and OE are placed side by side, but 2D integration uses an organic substrate with further distance between the OE and host whereas 2.5D has higher bandwidth density and energy efficiency.  2D approaches can result in larger packages and greater beachfront expansion as traces fan-out from the host to OEs.  The losses associated with the beachfront expansion translate into increased energy consumption on the host SerDes.
In practice Table~\ref{tab:legacy} shows that CPO with large beachfront expansion does not deliver substantially different energy efficiency than linear pluggable optics, when accounting for the host based SerDes.

\begin{figure}
    \centering
    \includegraphics[width=0.95\linewidth]{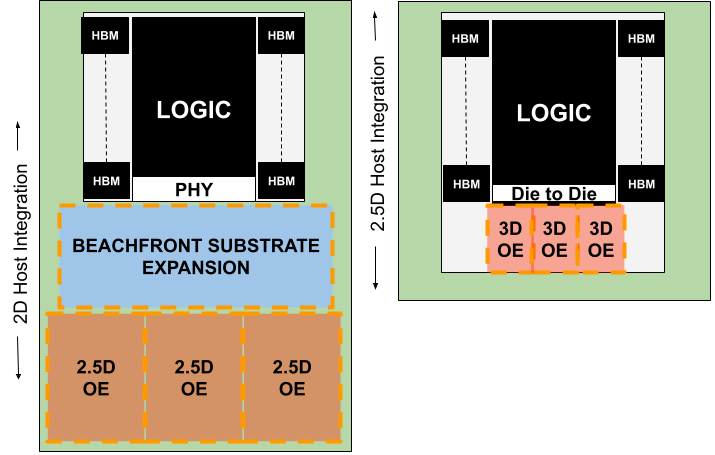}
    \caption{Difference between 2D and 2.5D integration of optical engines (OEs).  The left most approach shows larger 2.5D OEs with 2D host integration over an organic substrate and the resulting beachfront expansion.  The rightmost approach shows smaller 3D OEs that are 2.5D-integrated in close proximity to the host on an interposer or bridge.}
    \label{fig:enter-label}
\end{figure}

The optical engine itself may also be constructed in a variety of ways. 2D OEs lay out the electrical I/O and photonic components in a single plane rather than stacking a separate Electrical Integrated Circuit (EIC) and Photonic Integrated Circuit (PIC).  This approach requires more area and is shoreline limited with respect to the number of electrical interfaces it can support.  In a 2.5D approach the EIC and PIC may be stacked on top of each other, but have limited ability to pass power and signals from the substrate through the bottom die.  This requires routing those traces around the chip using redistribution layers and through mold vias rather than through the PIC or EIC, resulting in a larger OE.

In the next section we discuss a different approach taken by Passage, a fully 3D design, where EIC and PIC are stacked with Passage supporting power and signal delivery through the PIC itself with TSVs.  This allows for maximal design flexibility with respect to the placement of I/O, the lowest pJ/bit and highest and bandwidth density.

\section{Passage}
\label{sec:Passage}
\begin{figure}
    \centering
    \includegraphics[width=0.95\linewidth]{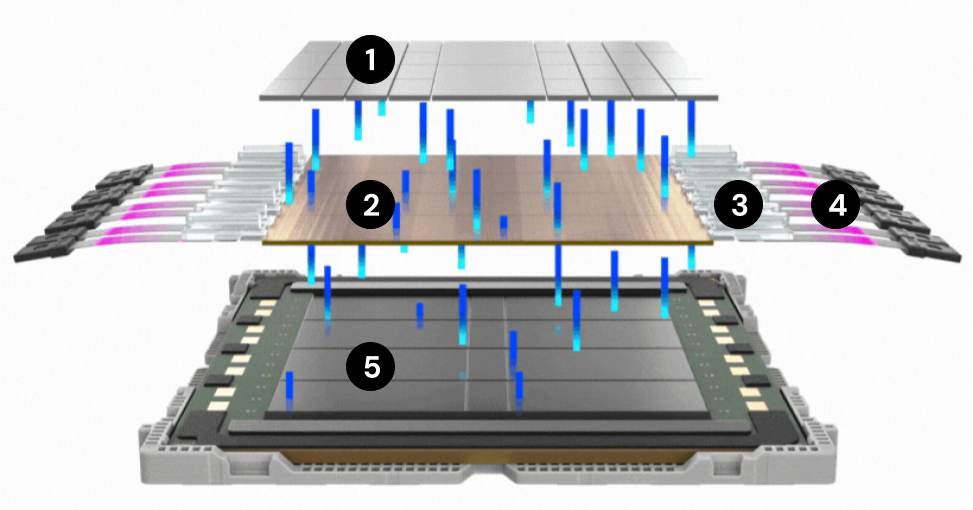}
    \caption{Exploded view of a Passage Interposer Solution: (1) EIC, (2) PIC, (3) Fiber Attach Unit, (4) fibers and (5) substrate.  Blue vertical lines distributed throughout package area represent I/O enabled without SerDes shoreline constraints.}
    \label{fig:PassageExplode}
\end{figure}

Lightmatter Passage is a 3D photonics platform.  3D stacking of electrical and optical interfaces places the optical elements directly underneath the footprint of electrical SerDes.  
This creates a tightly integrated, high-bandwidth, and low energy solution.  The energy efficiency of current Passage products is 2.3 pJ/bit for PIC and laser~\cite{lminterconnect25} plus SerDes, which is design dependent.  For short reach SerDes (e.g. XSR or VSR) this may be 1 pJ/bit~\cite{Davide2022} at 112 Gb/s PAM-4 or 2 pJ/bit with NRZ modulation.  This results in substantially greater efficiency than competing solutions (4.3 pJ/bit PIC, EIC, Laser and SerDes) -- and is lower than an electrical solution with DSP-based SerDes.

Passage is offered as either (1) 3D OE with 2.5D integration or (2) an optical interposer that sits under the entirety of the processor or switch.  An OE-based design is compatible with a variety of host designs provided they share a compatible die to die interface.  A Passage OE is similar in concept to HBM technology -- a set of 3D stacked dies, 2.5D-integrated with the host.  The die to die interface of an OE chiplet adds a small amount of power (0.5 pJ/bit\cite{UCIe}).  An interposer design offers compelling advantages such as cross-reticle waveguide stitching to support larger multi-reticle or waferscale designs.
To explain this in greater detail, Figure~\ref{fig:PassageExplode} shows an exploded view of a Passage Interposer design, where an EIC (1) sits on top of the PIC (2).  The EIC could be any computing device, but typical bandwidth hungry devices would be an GPU or a switch.  The EIC consists of one or more reticles and can be as large as a waferscale.  As discussed in Sec.~\ref{sec:background}, a traditional EIC places I/O and SerDes along its perimeter, whereas a Passage-enabled EIC can utilize I/O from anywhere within the chip area as indicated by vertical blue bars (signals) in the image.  The Passage PIC (2) is a combination of SiPh and conventional CMOS technology.  In addition to enabling optics, the PIC contains Through-Silicon Vias (TSVs) to provide the EIC with power and signaling from the Substrate (5). The PIC integrates all the components necessary to convert electrical signals to optical signals.  

\paragraph{Passage Modulators and Wavelength Division Multiplexing}
Passage uses arrays of Microring Modulators (MRMs) to support high-bandwidth wavelength division multiplexing (WDM).  MRMs are thermally controlled to resonate at different frequencies allowing for multiple wavelengths (also referred to as lambdas or colors) of light to share a single silicon waveguide or fiber.  Passage supports up to 16 colors per fiber, resulting in up to 1.792 Tb/s bandwidth per fiber at 112 Gb/s PAM-4.  This is 8 times higher density than CPO using single-lambda 224 Gb/s PAM-4 per fiber~\cite{BaillyHotChips}.  Alternatively, the WDM can utilize lower data rate SerDes for higher energy efficiency (such as 56 Gb/s NRZ).  Data transmission can even be bidirectional where TX and RX signals share the same fiber to improve fiber utilization.  Using WDM provides significantly greater bandwidth per fiber than single lambda approaches.
 %add a reference to Infiniband Coupe CPO

\paragraph{Datapath in Passage}
\begin{figure*}
    \centering
    \includegraphics[width=0.9\linewidth]{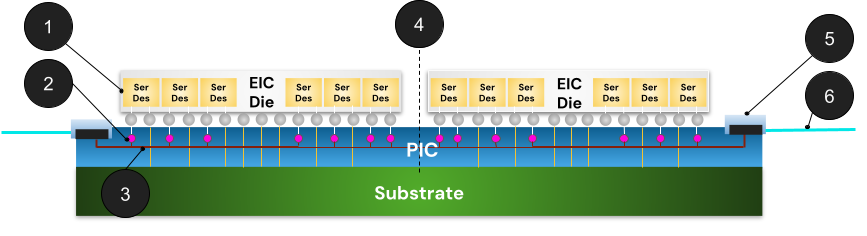}
    \caption{Illustration of Passage cut-through describing path of data through Passage.  Data is transmitted from multiple rows of SerDes distributed throughout the area of the EIC (1) to the Microring Modulator (MRM) within the PIC (2).  From the MRM arrays, multiple wavelengths of light travel along silicon waveguides (3).  In the case of a multi-reticle Passage design, cross-reticle waveguide stitching (4) creates a continuous path across the EIC reticle boundaries.  If the destination is a remote node, the waveguides egress through the Fiber Attach Unit (FAU) and transition to larger optical fibers (5).  The receive side path mirrors this process but includes a photodetector and transimpedance amplifiers.}
    \label{fig:cutthrough}
\end{figure*}

Figure~\ref{fig:cutthrough} shows an alternate view of the Passage design.  It highlights multiple rows of SerDes modules (1) throughout the area of the EIC.  The optical and electrical components of the Passage PIC (MRM, driver, waveguides, and transimpedance amplifier (TIA)) sit within the shadow of the EIC.   The stacked EIC and PIC design creates efficient use of area and maximizes the bandwidth per square mm.  The distance between the SerDes and the optical conversion (2) is under 100 $\mu$m, enabling the use of energy-efficient short reach SerDes without requiring DSPs.  Waveguides (3) allow optical transmission through silicon to another reticle within the same package or to FAUs.

\paragraph{Waveguide Routing and Optical Circuit Switching (OCS)} Waveguides provide flexible routing through the silicon, capable of bends, crossings and solid-state switching.  Within Passage Mach-Zender Interferometers (MZIs) enable 2 $\times$ 2 switching elements that are programmable and reconfigurable.  This creates an OCS capability within the GPU or Packet Switch host itself.  The OCS allows for (1) component-level resiliency to be built into the device, (2) multi-reticle designs, and (3) application-level optimizations via intra- and inter-Passage topology reconfigurations.  For waferscale designs Passage has demonstrated cross-reticle waveguide stitching, enabling a direct path from the fiber at the edge of the chip to any reticle on the device.  This is a key enabler for fully 3D devices.  

\paragraph{External Laser} Another benefit of Passage compared to pluggable modules and LPOs is the use of an external laser module.  The light generated by the laser is brought into Passage from a dedicated set of laser fibers before being split and directed to specific channels.  External lasers provide the ability to place the laser module where it is easier to control for thermal variability and stress, but more importantly, it allows the laser module to be replaced as a standalone unit.  This is crucial when the photonics are integrated into expensive packages such as a GPU.  Another benefit of external lasers is that the power consumption is out of package, which allows for greater power delivery to compute resources.

\section{System Design with Passage}
\label{sec:System}
In this section we examine three different approaches to constructing an optical GPU solution that enable a 512 GPU package (2048 GPU die) scale-up Pod.  The approaches are (1) LPO, (2) 2.5D CPO with 2D integration and (3) Passage optical interposer .  For each approach we generate projections of the power and energy required as well as the growth in area (package and board).   We assume a Single Layer of Switches (SLS) topology as explained in Section~\ref{sec:background}, such that each switch has at least one port connected to every GPU in the pod.

\paragraph{Port Definition}
We assume 448 Gb/s raw bandwidth per port which is the expected path of scale-up standards such as UALink~\cite{ualink}.  Larger port designs make it challenging to build high radix switches as the aggregate bandwidth within the switch fabric increases.  Smaller port designs lead to inefficient use of data fibers and poor bandwidth density. A 400 Gb/s port can be constructed differently dependent on the SerDes speed and number of lanes per port.  For Passage this is 8 lambda at 56 Gb/s NRZ encoding.  For other approaches this could be 4 lanes of 112 Gb/s PAM-4, or likely 2 lanes of 224 Gb/s PAM-4.  For scenarios where we assume a dense module (e.g. 1.6T DR8 LPO) a 400 Gb/s port requires breakout cabling to bifurcate the links so that the 400 Gb/s port can act as a distinct rail from GPU to switch in the SLS topology.

\subsection{Energy Efficiency}
\label{sec:energy}
%\begin{table}[h]
%    \centering
%    \begin{tabular}{|l|c|c|c|c|}
%        \hline
%        & LPO & 2.5D & 3D & Optical \\
%        & & CPO & CPO & Interposer \\
%        \hline
%        Total pJ/bit (Optics, SerDes, Laser, D2D) & 13 & 12 & 4.8 & \textbf{4.3} \\
%        \hline
%        In-package pJ/bit & 5 & 9.7 & 3.7 & \textbf{3.2} \\
%        \hline
%        Out-of-package pJ/bit & 8 & 2.3 & 1.1 & 1.1 \\
%        \hline
%    \end{tabular}
%    \caption{Comparison of Various Metrics Across Different Technologies \textbf{Internal Note: 2.3 pJ/bit M1000 and L200 PIC + Laser cited.  XSR 56 Gb/s NRZ assumes 2 pJ/bit. 0.5 pJ/bit for die to die.  1 pJ/bit assumed if bidirectional.}}
%    \label{tab:techenergycomp}
%\end{table}

\begin{table}[h]
    \centering
     \resizebox{\columnwidth}{!}{%  % scales to column width
    \begin{tabular}{|l|c|c|c|}
        \hline
        & 1.6T DR8 LPO & 224G 2.5D & 56GX8$\lambda$ Passage \\
        & 224G/lane & CPO & Interposer \\
        \hline
        In-package pJ/bit & 5 & 9.7 & \textbf{3.2} \\
        \hline
        Off-package pJ/bit & 8 & 2.3 & \textbf{1.1} \\
        \hline
        Total pJ/bit  & \raisebox{-2ex}{13} & \raisebox{-2ex}{12} & \raisebox{-2ex}{\textbf{4.3}} \\
        (Optics, Phy, Laser) & & & \\
        \hline
    \end{tabular}
    }
    \vspace{1em}
    \caption{Energy efficiency of (1) 1.6T 224G DR8 LPO, (2) 2.5D CPO with 2D integration, and Passage interposer design.  Host could be GPU, Switch or similar device.}
    \label{tab:techenergycomp}
\end{table}

In Table~\ref{tab:techenergycomp} we highlight the energy efficiency of an LPO 2.5D CPO and Passage Interposer design.

\paragraph{SerDes estimate}
We assume that both the LPO module and the 2.5D CPO with 2D integration are directly driven by host based SerDes.  Existing 112G-LR SerDes provide estimates of 4.5-6 pJ/bit\cite{shivnaraine202111, guo2022112}. At 224G speeds there are fewer results of measured energy efficiency.  Researchers at Synopsys published a 224Gb/s 3 pJ/bit 40 dB insertion loss design~\cite{10716779}, but this did not include the power dedicated to the DSP, which contributes significant additional power.  For these reasons 5 pJ/bit is our assumed energy efficiency for 224G-LR SerDes.  The SerDes power is included as part of the in-package power for GPU and switch estimates in this section.

\paragraph{LPO}
For a DR8 class (500m reach) SiPh LPO device we see a range of power numbers given in literature (6.25 pJ/bit \cite{HotI24Keynote} to 10-11.25 pJ/bit 800G (112G PAM-4)~\cite{EoptolinkLPO, LPOHisense} in existing devices).  OIF's estimates suggest that LPO devices could provide up to a 50\% savings in energy efficiency over traditional pluggable modules~\cite{OIFOFC}.  We use 8 pJ/bit for our estimates of a 1.6T DR8 module.

\paragraph{2.5D, 2D integrated CPO}
We use the data from the 2024 HotChips presentation of the Bailly CPO architecture~\cite{BaillyHotChips} as a reference point for a 2.5D Optical Engine.  In this presentation, a 51.2 Tb/s switch design results in 241 Watts of power for optical engines and 118 Watts of power for external lasers.  This is equivalent to 4.7 pJ/bit and 2.3 pJ/bit, respectively.  We assume the same energy efficiency for PIC and laser in a 224G design.  In ~\cite{BaillyHotChips} the 112G SerDes Phy are located on the host and must drive a signal over a large beachfront distance.  We assume the same 5 pJ/bit used in the LPO power estimates.  If the SerDes I/O die is connected to the host die over an 2.5D die-to-die interconnect this would add another 0.5 pJ/bit~\cite{UCIe}, but we assume a monolithic host with SerDes.

\paragraph{Passage}
For a Passage based optical interposer design we use the 2.3 pJ/bit number provided at ~\cite{lminterconnect25} for PIC and laser.  We further split this into 1.1 pJ/bit for the laser (off-package power) and 1.2 pJ/bit for the PIC (in-package power). For the SerDes we use 1 pJ/bit given by Tonietto~\cite{Davide2022} for a 112G PAM-4 XSR design and conservatively double that to 2 pJ/bit for 56G NRZ.  Passage is able to utilize much lower power SerDes due to the short drive distance required (less than 100 $\mu$m).

\subsection{Area Estimates}
\label{sec:area}

\begin{figure}
    \centering
    \includegraphics[width=0.95\linewidth]{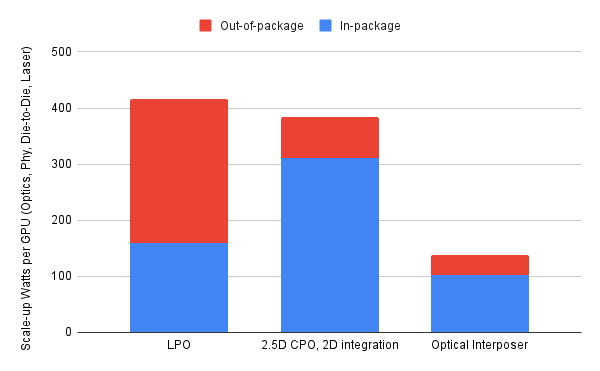}
    \caption{2.8$\times$ less power of Passage interposer over conventional optics for a 32 Tb/s unidirectional GPU.  Calculations based on values from Table~\ref{tab:techenergycomp}.}
    \label{fig:gpuscaleuppower}
\end{figure}

\paragraph{LPO}
 We use the specified 105.8 mm $\times$ 22.58 mm dimensions~\cite{osfpxd} for a total area of 2,389 sqmm per module.  We assume up to 16 channels (32 fibers) within a single extra dense module.  For a 3.2T module this results in an areal bandwidth density of 1.3 Gb/s/sqmm.

\paragraph{2.5D, 2D integrated CPO}
For this analysis we assume a 15 mm $\times$ 25 mm footprint for an optical engine with 10 mm of beachfront and 12.8 Tbps unidirectional bandwidth (using 224G SerDes).  This is reasonable given estimates of "roughly 1 Tbps/mm"~\cite{BrcmSemicon} and industry roadmaps~\cite{BaillyHotChips}.  This suggests areal density of approximately 34 Gb/s/sqmm or 24 Gb/s/sqmm when accounting for beachfront.

\paragraph{Passage}
An optical interposer design sits under the host monolithic or multi-chip module.  There is a small amount of area expansion typically to account for fiber attach mechanisms.  We use 5 mm of Passage expansion beyond the host chip.  The other dependency is the number of fibers being attached. These fibers are 127 $\mu$m and can be estimated at 4 fibers per mm of shoreline.  For a 56G 8$\lambda$ design this means two TX and two RX fibers per 5 sqmm or 160 Gb/s/sqmm.  This is particular to this design point as some Passage designs may (1) interleave TX and RX within the same fiber to increase this density and (2) utilize 112G PAM-4 modulation.  \textbf{For a 400 Gb/s port definition, this represents a 123$\times$ and 6.6$\times$ reduction in additional optical area compared to LPO and 2.5D/2D-integrated CPO, respectively}

\subsection{Impact on GPU and Switch Design}
\paragraph{GPU}
GPU Packages continue to deliver 2-fold aggregate performance increases per generation. Much of these gains come from increases to package size and the number of GPU and memory dies with a modest 15\% improvement in performance due to increases in process (e.g. N7 to N5 process with equivalent power) ~\cite{TSMC7to5}.  %Alternatively, advancements in process can provide up to 30\% decrease in power consumption at the same performance of the previous generation.
In the 2028 timeframe, high-end GPUs will consist of 4 logic dies with stacks of HBM on two sides of package perimeter.  The logic dies are configured in a 2X2 or 1X4 configuration.  We assume a full reticle is 26 mm x 33 mm and that stacks of HBM are 13 X 11 mm.

Recent extensions of roadmaps~\cite{gtcroadmap} show a 2027-28 GPU in a configuration similar to Figure~\ref{fig:4x1} with 16 stacks of HBM4 (north and south sides) totaling 209 Tb/s (26 TB/s) of memory bandwidth (6.4 GT/s).  This leaves two sides of the package available for I/O.  For I/O we assume 32 Tb/s RX and 32 Tb/s TX bandwidth which provides a ratio of 6.67:1 of HBM to scale-up bandwidth per GPU.

Achieving 32 Tb/s of bandwidth on a GPU would require 160 channels of 8$\times$400 Gb/s (224 Gb/s-PAM4).  This is equivalent to 10 OSFP-XD modules.  In aggregate this is over 20,000 sqmm of board area. The bandwidth required per device would likely lead to the use of co-packaged copper or copper flyover cables from the host to the modules to reduce PCB losses.

\begin{figure}
    \centering
    \includegraphics[width=0.95\linewidth]{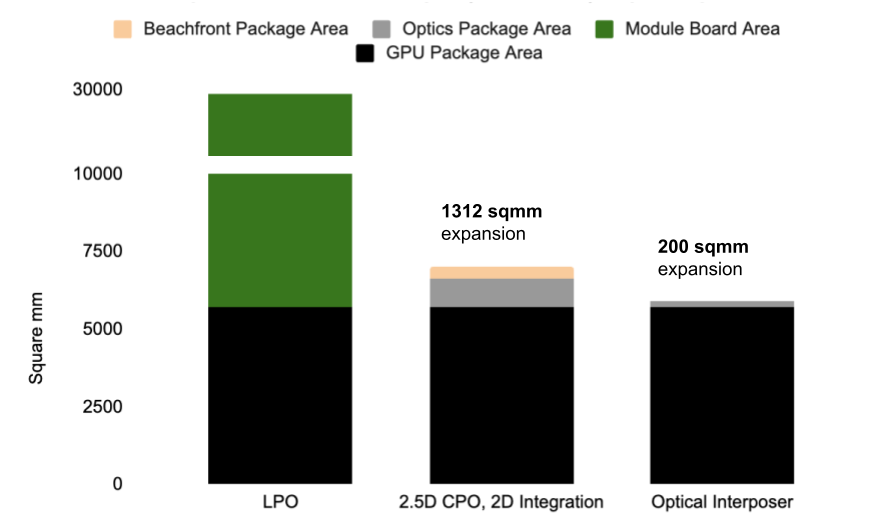}
    \caption{Comparison of the area required to support 32 Tb/s unidirectional bandwidth on a four reticle GPU.  Includes GPU package (logic and HBM), optics on-package, package beachfront expansion, and board expansion. }
    \label{fig:areacomparegpu}
\end{figure}

For a 2.5D CPO solution, this would require 3 12.8T OEs, but using the areal bandwidth densities previously calculated, this would result in ~1312 mm of combined OE plus beachfront expansion.  For a Passage interposer design, this is a relatively small 200 sqmm.
Figure~\ref{fig:areacomparegpu} shows LPO modules require a massive area of real estate on the board compared to co-packaged optics and interposer based designs. The CPO solution results in a 23\% increase in package area of the GPU compared to a 3.5\% increase for a optical interposer.

\paragraph{Switch}

For SLS topologies, the design point is a 200 Tb/s switch package (229 Tb/s raw bandwidth) with 512 ports.  We expect for the switch fabric of these designs to be multi-reticle based on area required for memory, NoC and SerDes.  For a switch fabric design using LPO or CPO the main constraint is shoreline available for SerDes.  This requires enough shoreline to place 128 $\times$8-224G SerDes macros.  Assuming aggressive 1.5D stacking of SerDes and 3 mm of shoreline per macro would result in 256 mm of required shoreline.  Unfortunately, reticle size limits (33$\times$26 mm) prevent this from fitting on the combined edges of two full reticles.  LPO and CPO could require a 4 reticle design for this amount of bandwidth.  Alternatively, \textbf{Passage  provides tremendous benefits to reducing the total package area by distributing the SerDes throughout the fabric die, rather than the shoreline.}
From the perspective of pJ/bit, the values in Table~\ref{tab:techenergycomp} are identical.  Accounting for the 200Tb/s per switch \textbf{Passage results in 1.5KW of power savings per switch package.}  

\section{Application Modeling}
\label{sec:App}
\subsection{Analytical modeling tool}
To evaluate performance, we developed an analytical performance modeling tool for LLMs that enables rapid evaluation of different architectures and deployment strategies without the need for actual implementation or empirical testing. The tool models execution time as a combination of computation, memory access, and communication costs, expressing each component through analytical formulas that capture the key characteristics of LLM training \cite{10707191}. Similar approaches have been developed for general LLM modeling \cite{10.1145/3581784.3607102, 10609708} and specifically focusing on MoE architectures \cite{10.1145/3503221.3508418}.

Our methodology decomposes LLM execution into its constituent operations - including attention computation, feed-forward networks, and in the case of MoE, expert routing. For each operation, we implement analytical expressions that account for hardware capabilities (like compute throughput and memory bandwidth), system topology (including high-speed interconnects and slower inter-node networks) and various parallelization strategies. The tool provides  modeling of key parallelization strategies for LLM training, including data parallelism (DP), tensor parallelism (TP), pipeline parallelism (PP), and expert parallelism (EP). This analytical approach allows us to model how different architectural choices and system configurations affect overall performance.

We model collective communication operations using the widely-adopted Hockney model \cite{HOCKNEY1994389}. This model expresses the time for a communication operation as $\alpha + \beta n$, where $\alpha$ represents the latency (startup time), $\beta$ is the transfer time per byte, and $n$ is the message size in bytes. This simple yet effective model captures both the fixed overhead of initiating communication and the bandwidth-dependent cost of data transfer. We implemented analytical models for key collective operations used in distributed LLM training including all-gather, reduce-scatter, all-reduce, and all-to-all operations on various topologies.

%\begin{figure}
%        \centering
%    \includegraphics[width=0.90\linewidth]{figs/one_expert_per_dp_rank_v3.png}
%    \caption{One expert per DP rank / TP group} 
%    \label{fig:1_expert_per_rank}
%\end{figure}
%\begin{figure}
%        \centering
%    \includegraphics[width=0.90\linewidth]{figs/two_experts_per_dp_rank_v3.png}
%   \caption{Multiple expert per DP rank / TP group} 
%    \label{fig:multi_expert_per_rank}
%\end{figure}

\begin{figure}
    \centering
    \begin{subfigure}[b]{0.48\textwidth}
        \centering
        \includegraphics[width=\linewidth]{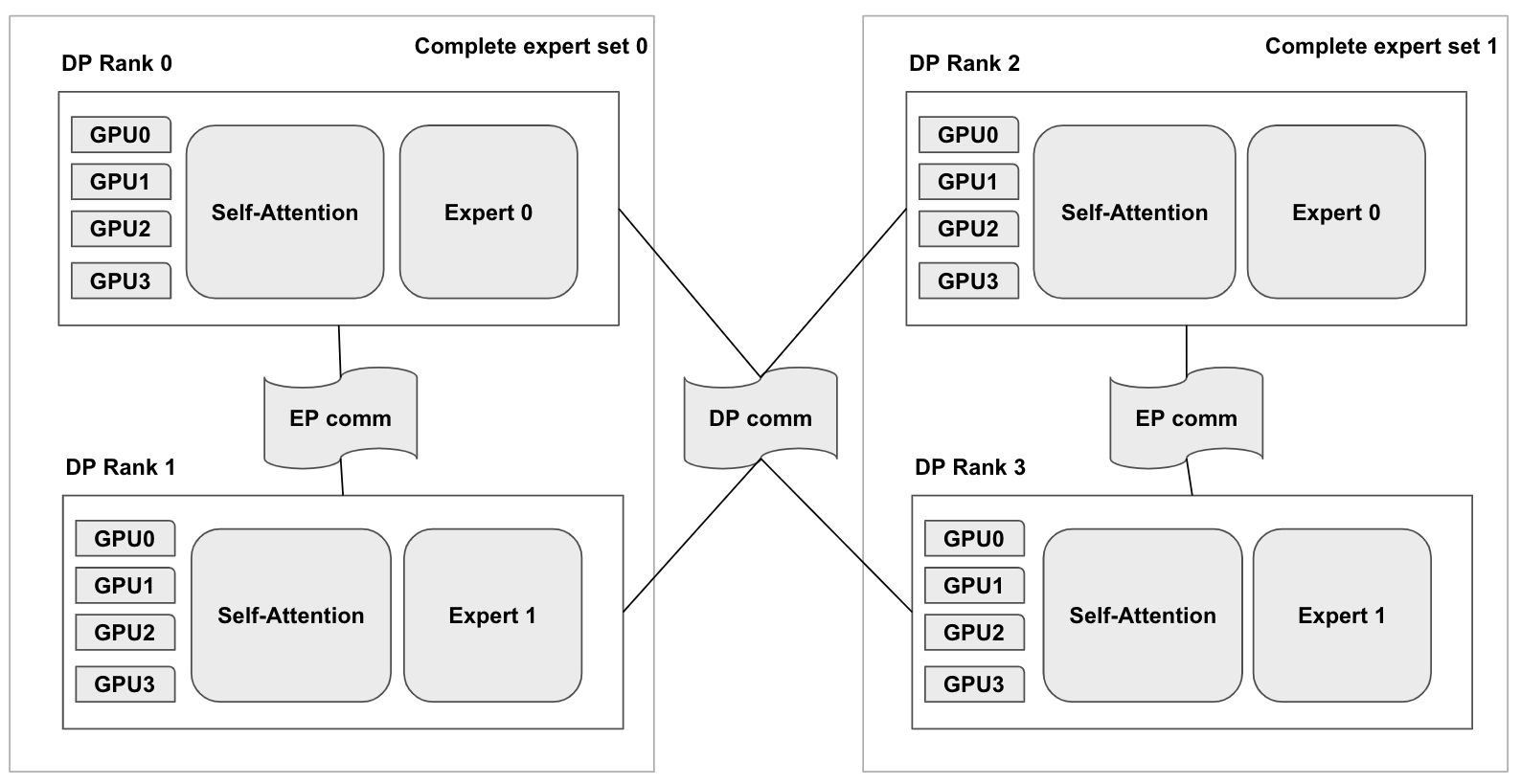}
        \caption{One expert per DP rank / TP group}
        \label{fig:1_expert_per_rank}
    \end{subfigure}
    \hfill
    \begin{subfigure}[b]{0.48\textwidth}
        \centering
        \includegraphics[width=\linewidth]{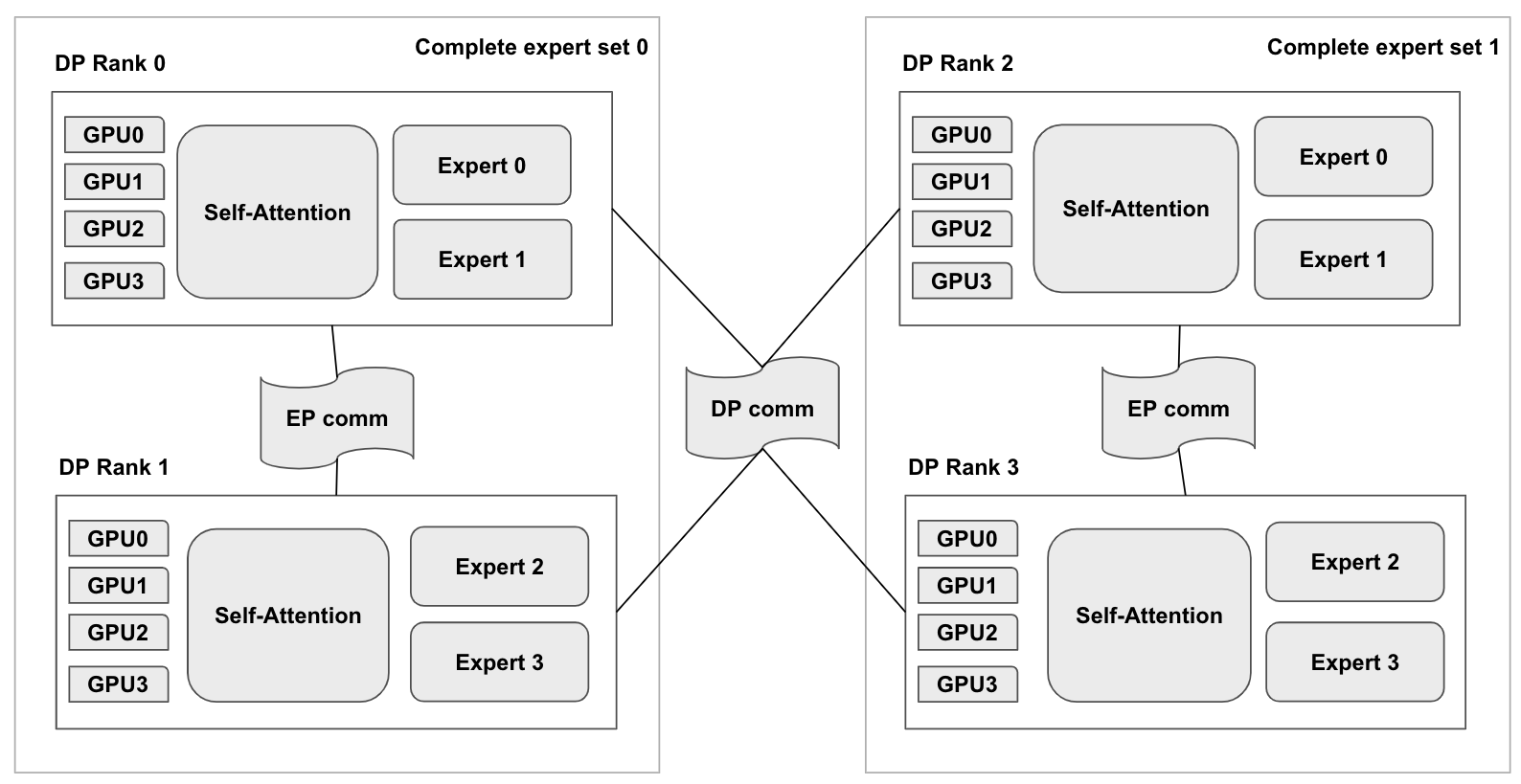}
        \caption{Multiple experts per DP rank / TP group}
        \label{fig:multi_expert_per_rank}
    \end{subfigure}
    \caption{Expert distribution strategies across DP ranks}
    \label{fig:expert_distribution}
\end{figure}

\subsection{Mixture-of-Experts workload scenarios}
In MoE models, we increase the model parameters at a lower compute cost than a similarly sized dense model by selectively activating experts. In each layer, the attention blocks are shared by all experts, with tokens then routed to a subset of activated experts by a small linear model.  Each expert is an identically dimensioned feedforward network taking the place of the single FFN in a dense transformer layer (Figure \ref{fig:moe_layer}). In the dense transformer case, tensor parallelism is used to partition the single FFN computation across multiple GPUs. In the MoE case, we now need an instance of the FFN for each expert in a given layer, and communications between each expert and its corresponding routing layer (Figure~\ref{fig:1_expert_per_rank}). Typically, the entire available high-bandwidth domain is allocated to the tensor parallel group, and the expert parallelism communications goes over a slower path such as Ethernet or other data center networking. 

Passage has both higher aggregate bandwidth and higher radix, allowing expert parallel communications to move from the slower network onto the high-bandwidth domain. At typical tensor parallel sizes of 8 or 16 nodes, this leaves room for up to 64 full size experts or a larger number of smaller experts, which is often preferable\cite{tang2025panguultramoetrain}. 

As discussed in the next section, we organize expert parallelism by allowing each DP rank to host multiple experts, with the original TP group subdivided into several expert TP groups - one for each expert in the DP rank (Figure~\ref{fig:multi_expert_per_rank}). Following optimizations from \cite{10.1145/3577193.3593704}, we eliminate redundant token transfers in this hybrid scheme. The presence of experts also modifies traditional DP communication patterns. With sufficient DP ranks, multiple complete sets of experts exist in the system, where each complete set contains exactly one instance of every unique expert required to process any possible routing decision. Gradient synchronization occurs selectively between corresponding expert copies located in different complete expert sets, rather than across all DP ranks uniformly as it is done for the attention part.

\subsection{Scaling MoE architectures: expert count and fine-grained segmentation}

MoE architectures show a clear evolution in expert scaling and activation patterns. Early models like Switch Transformers \cite{fedus2022switchtransformersscalingtrillion} demonstrated the potential of sparse architectures with 64 experts and single expert activation per token. OLMoE \cite{muennighoff2025olmoeopenmixtureofexpertslanguage} maintained the same expert count but increased activation to 8 experts per token, showing the benefits of combining multiple expert outputs. This trend toward higher expert counts continues with DeepSeek-V3 \cite{deepseekai2025deepseekv3technicalreport} and Pangu Ultra MoE \cite{tang2025panguultramoetrain}, both employing 256 experts while maintaining 8 expert activations per token. Notably, Pangu Ultra MoE's ablation studies suggest diminishing returns beyond 256 experts, indicating a sweet spot for balancing performance and computational efficiency.

This trend toward more experts is well-justified by the increased modeling capacity and flexibility it provides. With more experts and higher expert activation counts per token, the model can develop more specialized capabilities and combine them more effectively. While early MoE models like Switch Transformer activated only one expert per token, modern architectures activate multiple experts from a larger expert pool, enabling more sophisticated compositions of specialized knowledge. This combination of increased expert count and multiple expert activations per token allows models to leverage several specialists simultaneously while maintaining narrow, focused expertise within each expert.

To make these larger expert pools computationally feasible, fine-grained expert segmentation \cite{krajewski2024scalinglawsfinegrainedmixture,dai2024deepseekmoeultimateexpertspecialization} has emerged as a crucial technique. The key insight is to partition the hidden dimension of each expert's feed-forward layer - if the original expert had a hidden dimension of size $d_{\mathit{ff}}$ (typically $4  d_{\mathit{model}}$), each fine-grained expert now operates on a smaller hidden dimension of size $d_{\mathit{ff}}/m$, where $m$ is the number of fine-grained experts created from each original expert. By activating $m$ times more experts per token while reducing each expert's hidden dimension by a factor of $m$, this approach maintains constant computational costs while enabling the benefits of larger expert pools - effectively enabling access to sophisticated MoE architectures that would otherwise be computationally prohibitive.

\section{Results}
\label{sec:Results}
We evaluate different MoE configurations in the context of a large-scale language model training setup. The base architecture is a 120-layer decoder-only transformer with model dimension ($d_{\mathit{model}}$) 12288 and 128 attention heads, following the GPT family of models. The model employs Megatron-style tensor parallelism \cite{shoeybi2020megatronlmtrainingmultibillionparameter} for both attention and feed-forward computations. The total parameter count of such model is 4.7T.

The training configuration maintains consistent parallelization dimensions across all scenarios: tensor parallelism degree of 16, data parallelism degree of 256, and pipeline parallelism degree of 8, running on a fixed cluster size of 32,768 GPUs. Each GPU delivers 8.5 PFlops of compute performance using BF16 precision. Each Ethernet link provides 1600 Gb/s of unidirectional bandwidth. The training processes a global batch size of 4096 with sequence length 8192, targeting 13T tokens of training data. 

We evaluate these configurations across two distinct network scenarios:
\begin{itemize}
    \item A network with a scale-up pod size of 144 GPU packages and 14.4 Tb/s unidirectional bandwidth per GPU, representing the limits of electrical scale-up solutions.
    \item Passage: An optical network with a scale-up pod size of 512 GPU packages and 32 Tb/s unidirectional bandwidth per GPU.
\end{itemize}

Within these fixed infrastructure constraints, we explore different MoE scaling strategies as shown in Table \ref{tab:cluster-config}. The expert granularity parameter m shows how each configuration implements fine-grained experts. Starting with m = 1 in Config~1 (standard experts with full $d_{\mathit{ff}}$ hidden dimension), each subsequent configuration splits the experts into progressively smaller units. For instance, Config 4 with m = 8 divides each original expert into 8 fine-grained experts, each with a hidden dimension of $d_{\mathit{ff}}/8$.

The distribution of experts across data parallel (DP) ranks follows the same progression. This arrangement ensures efficient communication patterns, as the number of experts per DP rank increases proportionally with the total expert count and granularity. This systematic scaling of both expert count and granularity allows us to evaluate how different expert configurations perform under realistic hardware and networking constraints typical of large-scale AI training clusters.
 
\begin{table}[h]
    \centering
     \resizebox{\columnwidth}{!}{
    \begin{tabular}{|l|c|c|c|c|}
        \hline
        \textbf{Parameter} & \textbf{Config 1} & \textbf{Config 2} & \textbf{Config 3} & \textbf{Config 4} \\
        \hline
        Active / total experts & 1/32 & 2/64 & 4/128 & 8/256  \\
        \hline
        Expert granularity (m) & 1 & 2 & 4 & 8  \\
        \hline
        Experts per DP rank & 1 & 2 & 4 & 8 \\
        \hline
    \end{tabular}
    }
    \vspace{1em}
    \caption{Cluster configuration parameters}
    \label{tab:cluster-config}
\end{table}

%\begin{figure}[htbp]
%   \centering
%   \includegraphics[width=\columnwidth]{figs/training_time_same_radix_v3.png}
%   \caption{Relative performance of both architectures assuming the same radix-512 (normalized to Config1 Passage baseline)}
%   \label{fig:resultssameradix}
% \end{figure}
%    \hfill
% \begin{figure}[htbp]
%        \centering
%        \includegraphics[width=\columnwidth]{figs/training_time_different_radix_v3.png}
%        \caption{Relative performance with system-specific radix settings: Passage (512) vs Alternative (144) (normalized to Config1 Passage baseline)}
%        \label{fig:resultssystemspecificradix}
% \end{figure}

\begin{figure}[htbp]
   \centering
   \includegraphics[width=0.95\columnwidth]{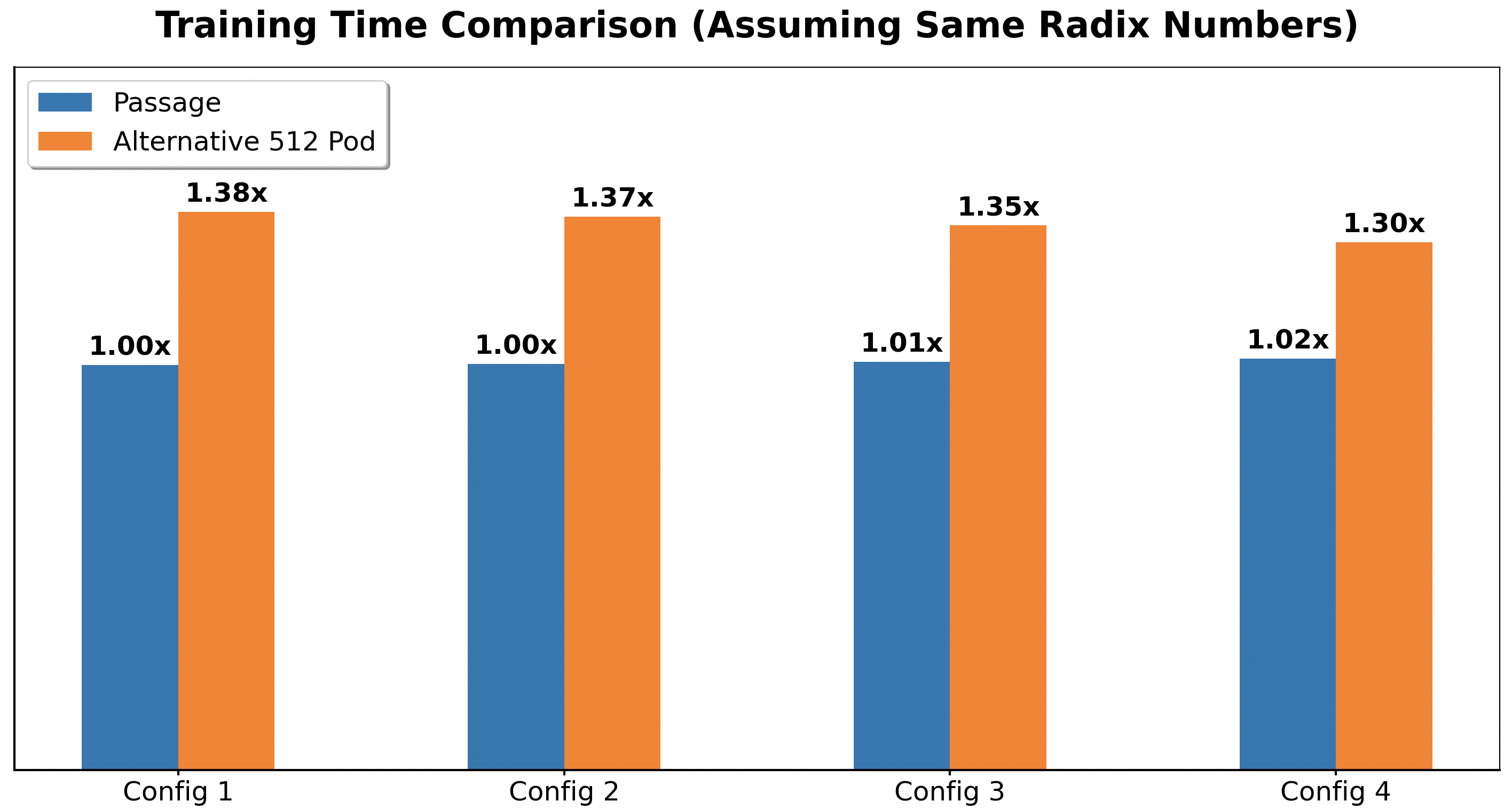}
   \caption{Relative performance of both architectures assuming the same radix-512 (normalized to Config~1 Passage baseline)}
   \label{fig:resultssameradix}
   \vspace{-0.3cm}  % Reduce space between figures
\end{figure}

\begin{figure}[htbp]
   \centering
   \includegraphics[width=0.95\columnwidth]{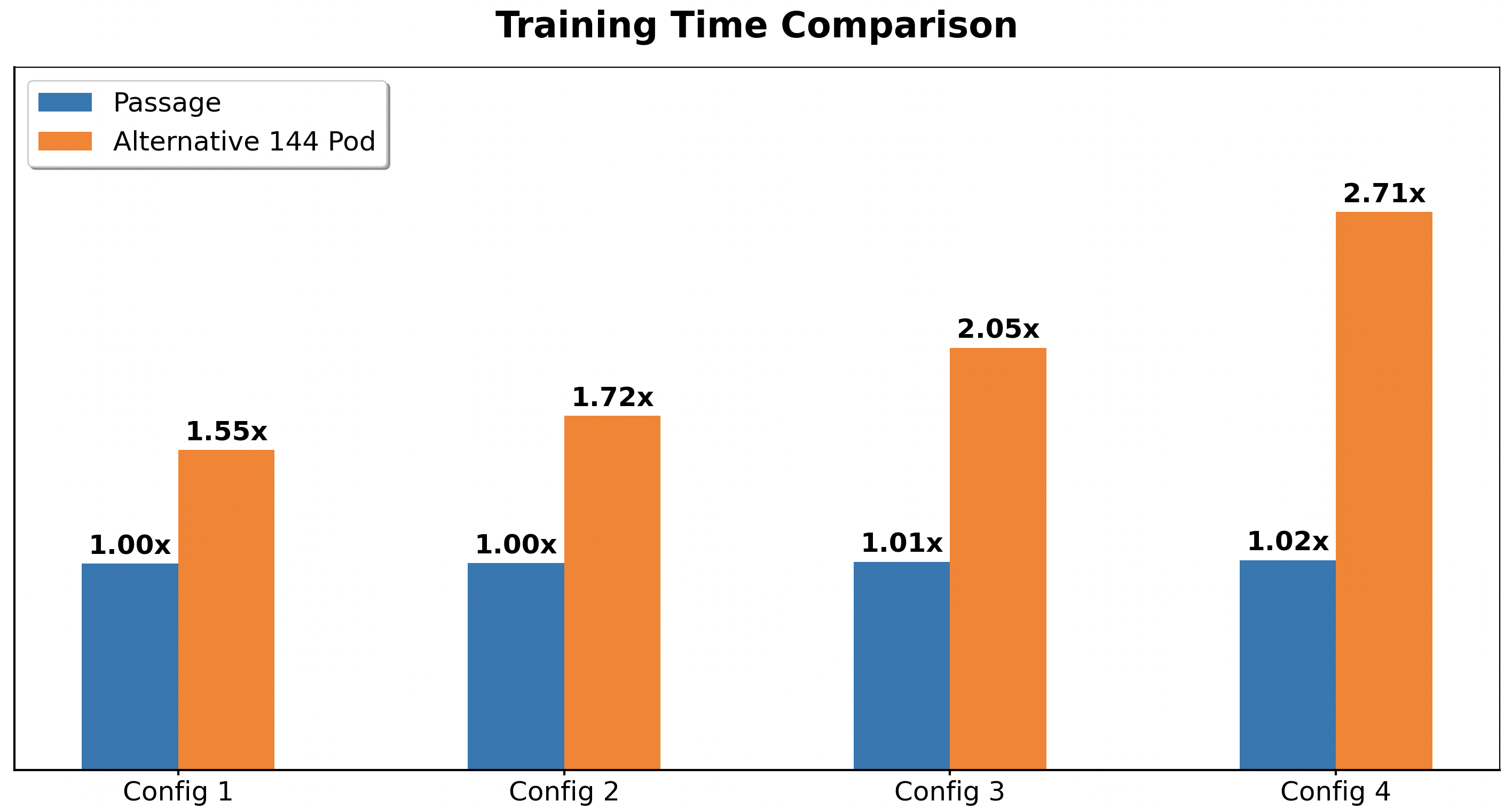}
   \caption{Relative performance with system-specific radix settings: Passage (512) vs Alternative (144) (normalized to Config~1 Passage baseline)}
   \label{fig:resultssystemspecificradix}
\end{figure}

We assume that tensor parallel groups are placed in the high bandwidth domain first, and expert parallel groups are placed in the high bandwidth domain if there is room to add them. Placing multiple smaller experts together can allow a larger number of experts to stay within the high bandwidth domain. In the Passage-based configuration, up to 512 GPU nodes can be placed in the high bandwidth domain. We assume a limit of 144 nodes for the alternate configuration. 

The performance comparison between Passage and the alternative solution reveals significant differences in how these architectures scale across different MoE configurations. Our analysis focuses on relative performance scaling, using Config 1 of Passage as the baseline reference point.

To isolate the impact of network bandwidth differences between architectures, we first compare both systems using pod sizes of 512 GPU packages (Figure~\ref{fig:resultssameradix}). Even with identical network topology, the higher bandwidth of Passage (32 Tb/s vs 14.4 Tb/s) demonstrates clear advantages in scaling efficiency. Passage shows minimal overhead as configurations become more complex, with Config 4 requiring only 1.02x the training time of Config 1. The alternative requires 1.4x longer training time compared to Passage for Configs 1 and 2, and 1.3x longer for Configs 3 and 4. The change in the alternative system's relative performance is explained by its communication bottleneck: as expert tensor parallelism distributes each expert across fewer GPUs in successive configurations while maintaining the same communication volume per GPU, the bandwidth pressure decreases. This highlights the significant impact of bandwidth differences even when network topologies are matched.

When comparing systems with their architecture-specific network configurations (512 GPU Pod at 32 Tb/s uni-directional for Passage vs 144 GPU Pod at 14.4 Tb/s uni-directional for the alternative), the performance gap widens substantially (Figure~\ref{fig:resultssystemspecificradix}). This divergence becomes particularly pronounced with finer-grained expert configurations, where both the total expert count and active experts per token increase (from 1/32 experts in Config 1 to 8/256 in Config 4). The alternative system requires 1.6x longer training time than Passage for Config 1, increasing to 2.7x for Config 4, while Passage scales efficiently. The combination of lower radix and bandwidth in the alternative system amplifies the communication bottlenecks from expert routing, resulting in significantly degraded scaling efficiency.

This scaling challenge manifests primarily through the expert all-to-all communication pattern, where tokens must be routed to their designated experts across the distributed system \cite{lepikhin2020gshardscalinggiantmodels}. With more fine-grained experts and higher activation counts per token, each input effectively requires more network traversals to accumulate its computational results. The alternative architecture, which relies more heavily on scale-out networking for expert communication, becomes increasingly bottlenecked by this growing communication volume.

Passage's architecture alleviates this pressure by maintaining experts within high-bandwidth domains. This architectural choice means that even as we scale to more fine-grained experts with higher activation counts, the critical expert communication patterns remain within high-bandwidth pathways. 
%This advantage is clearly reflected in the performance data, where Passage shows only a 1.51x slowdown in Config 4 %compared to the 5.19x deterioration seen in the alternative system.

This architectural efficiency has implications beyond pure
performance metrics.  Traditional MoE systems often require careful tuning of load balancing losses to prevent network congestion and ensure even expert utilization. 
For instance, \cite{deepseekai2024deepseekv2strongeconomicalefficient} uses device-limited routing restricting each token's experts to at most M devices. Passage's architecture keeps experts within high-bandwidth domains, eliminating strict routing constraints while maintaining stable performance at scale, thus simplifying training and enabling more flexible expert utilization.
%For instance,  \cite{deepseekai2024deepseekv2strongeconomicalefficient} employs device-limited routing that ensures each token's target experts are distributed on at most M devices, where M is a configuration parameter. Passage's architecture, by maintaining experts within high-bandwidth %domains, reduces the need for such strict routing constraints while maintaining stable performance as expert counts increase, simplifying the training process and allowing for more flexible expert utilization patterns.

% These results suggest that Passage's architecture provides superior scaling efficiency for large-scale MoE models. Its ability to maintain relatively stable performance across configurations, particularly when increasing expert counts and granularity, indicates better handling of communication patterns and network utilization. 

\section{Conclusions and Future Work}
\label{sec:Conclusions}

Our modeling demonstrates the profound impact of 3D integrated optics on the efficiency of MoE model training. The results show that the expanded radix and higher aggregate bandwidth of the 3D optical interconnect deliver substantial performance gains. When isolating bandwidth effects by comparing Passage against a hypothetical 512-radix version of the alternative system, the higher bandwidth alone delivers up to \textbf{1.4x speedup}. The performance gap widens further when comparing actual system configurations - Passage's 512-radix network versus the alternative's 144-radix topology. Here, Passage achieves a \textbf{2.7x speedup} for the most demanding configuration (Config 4) by accommodating more expert parallel communications within the high-bandwidth domain. Critically, Passage's elimination of communication bottlenecks ensures that additional compute capacity can be fully utilized rather than sitting idle waiting for data transfers - enabling higher computational intensity that would be wasted in bandwidth- and radix-constrained architectures. These findings underscore that MoE workloads effectively leverage Passage's expanded optical interconnect radix, accelerating traffic that would otherwise traverse slower scale-out networks. The combined benefits of higher bandwidth and connectivity enable Passage to maintain strong scaling efficiency even as expert counts and routing complexity increase.  Future work will further optimize 3D integrated optics technology to leverage the full potential of high-radix optical interconnects and optical circuit switching.

% \bibliographystyle{plainurl} % We choose the "plain" reference style
% \bibliography{refs} % Entries are in the refs.bib file

% begin try alt bib formatting
\begin{spacing}{0.9} % Slightly condensed line spacing within each entry
    \printbibliography
\end{spacing}
% end try alt bib formatting

\end{document}